\documentclass[authoryear, 12pt,1p]{elsarticle}
\usepackage{lipsum}
\makeatletter
\def\ps@pprintTitle{%
 \let\@oddhead\@empty
 \let\@evenhead\@empty
 \def\@oddfoot{}%
 \let\@evenfoot\@oddfoot}
\makeatother

\usepackage{graphicx}
\usepackage{color}
\usepackage{amsfonts}
\usepackage{amsmath}
\usepackage{amsthm}
\usepackage{array}
\usepackage{verbatim}
\usepackage{mathabx}
\usepackage{natbib}
\usepackage{sidecap} 
\usepackage{float}
\usepackage{overpic}
\usepackage{framed}
\usepackage{txfonts}

\begin{document}

\begin{frontmatter}

\title{A Simulation of the Effects of Receive Field Contrast on Motion-Corrected EPI Time Series}

\author[ucb]{D. Sheltraw \corref{cor2}}
\ead{sheltraw@berkeley.edu}

\author[ucb]{B. Inglis}

\cortext[cor2]{Principal corresponding author}

\address[ucb]{Henry H. Wheeler Jr. Brain Imaging Center, University of 
California, Berkeley, CA, 94720, USA}

\begin{abstract}The receive field of MRI imparts an image contrast which is
spatially fixed relative to the receive coil. If motion correction is used 
to correct subject motion occurring during an EPI time series then the 
receiver contrast will effectively move relative to the subject and produce
temporal modulations in the image amplitude. This effect, which we will call 
the RFC-MoCo effect, may have consequences in the analysis and interpretation 
of fMRI results. There are many potential causes of motion-related noise and 
systematic error in EPI time series and isolating the RFC-MoCo effect would
be difficult. Therefore, we have undertaken a simulation of this effect to 
better understand its severity. The simulations examine this effect for a 
receive-only single-channel 16-leg birdcage coil and a receive-only 12-channel 
phased array. In particular we study: (1) The effect size; (2) Its 
consequences to the temporal correlations between signals arising at different 
spatial locations (spatial-temporal correlations) as is often calculated in 
resting state fMRI analyses; and (3) Its impact on the temporal signal-to-noise
ratio of an EPI time series. We find that signal changes arising from the 
RFC-MoCo effect are likely to compete with BOLD (blood-oxygen-level-dependent) 
signal changes in the presence of significant motion, even under the assumption
of perfect motion correction. Consequently, we find that the RFC-MoCo effect 
may lead to spurious temporal correlations across the image space, and that 
temporal SNR may be degraded with increasing motion.
\end{abstract}

\end{frontmatter}

\section{Introduction}
\label{intro}

Subject motion can be a significant source of noise in BOLD 
(blood-oxygen-level-dependent \citep{bold}) functional magnetic resonance 
imaging (fMRI), a neuroscience research tool employing time series 
of gradient echo images, most commonly using the echo planar imaging (EPI) 
pulse sequence. This problem has been recognized since the 
inception of fMRI \citep{Hajnal}. The most obvious concern is that motion of 
the brain, with its contrast fixed relative to the head, will produce 
temporal variations at any given point in image space that can mask the BOLD 
or be erroneously attributed to the BOLD effect. But there can also be 
contrast within brain images that is fixed relative to the scanner. Such 
contrast can potentially be exascerbated by motion correction since following 
correction such contrast would then move relative to the brain and produce its 
own temporal variation. 

With the proliferation of MRI systems employing multichannel receiver arrays 
(phased arrays) arises greater potential sensitivity to motion-related error 
and noise in EPI time series data due to scanner-fixed receiver contrast. 
Although multichannel receiver arrays hold great promise for improved 
tSNR (temporal signal-to-noise ratio) it presently appears that motion-related 
noise may be limiting their usefulness \citep{Hartwig}. The effects of motion 
can be especially problematic when accelerated imaging \citep{griswold1, SENSE}
is used in EPI time series \citep{ACS_paper_2}. However, even in the absence of 
acceleration there are many potential motion-related effects that can prevent 
multichannel receiver arrays from achieving their motion-free tSNR potential. 
One such effect, arising from scanner-fixed contrast, which we will call the 
RFC-MoCo effect (Receive Field Contrast Motion Correction effect), may occur 
when motion correction is applied to EPI time series data possessing 
significant receive field contrast.

There are many potential causes of motion-related noise in EPI time series.
Here are some of the acknowledged important sources of noise and systematic 
error in non-accelerated EPI:

\begin{enumerate}
\item{Poor motion correction: Most motion correction algorithms used in fMRI
    \citep{AFNI, SPM2, FSL, AIR} assume that a volume of image slices moves as 
    a rigid body. Since a volume of data is usually collected over a time 
    interval of 1 to 2 seconds there is clearly enough time for the head to 
    move during such an interval, thereby violating the rigid volume assumption.
    This may lead to bad estimates of motion parameters and inadequate 
    correction of motion-related contrast fluctuations. Furthermore the issue 
    is often complicated by the use of additional temporal interpolation 
    \citep{Roche,Bannister} as well.}
\item{Spin history: The steady state of the pre-excitation magnetization, 
    established after a few TR (repetition time) periods in the absence of 
    motion, is perturbed by motion orthogonal to the imaging planes of 2D 
    multislice EPI \citep{Spin_hist}, potentially causing increased noise and 
    systematic effects upon spatial-temporal correlations.}
\item{Main field inhomogeneity: The presence of an object within the 
    strong main magnetic field $B_0$ of MRI leads to increased inhomogeneity 
    of the field. The inhomogeneity results in image distortions which increase
    in severity with field strength \citep{JB}. Shim coils are used to try to 
    restore field homogeneity but shim coils are fundamentally limited with 
    respect to the fields they can compensate. When the object moves the 
    inhomogeneity varies with respect to time which may potentially increase
    noise and systematic error in the time series images while decreasing 
    image-space signal strength.}
\item{Receive field distortion: A receiver coil loaded with a head will distort 
    the coil's receive field \citep{coil_load}. This distortion should be more 
    pronounced at high main field strengths \citep{high_field_review}. As the 
    head moves this distortion will vary in time thereby adding temporal 
    fluctuations in an EPI time series.}
\item{RFC-MoCo effect: The receive field, when not spatially homogeneous, will
    impose an image contrast that is fixed relative to the scanner
    \citep{Hartwig}. When an assumed perfect motion correction is applied to a 
    time series of such images the receive field effectively moves relative to 
    the imaged object. This temporally varying image contrast may potentially 
    produce tSNR degradation and systematic effects upon spatial-temporal 
    correlations.}
\item{TFC-MoCo effect: The transmit field, when not spatially homogeneous, 
    will also impose an image contrast that is fixed relative to the 
    scanner and will produce temporally varying image contrast in 
    a motion-corrected time series of images in a manner analogous to the 
    RFC-MoCo effect. Most commercial scanners used in fMRI research today
    make use of a large
    diameter birdcage body coil to produce the transmit field. The large 
    diameter, as compared to a head coil, increases the transmit field 
    homogeneity and therefore decreases the potential TFC-MoCo effect.}
\end{enumerate}

\noindent The RFC-MoCo effect is the focus of this paper. In particular, we 
seek to estimate the effect size of tSNR degradation and any spurious 
spatial-temporal correlations induced by the receive field contrast. 
This is an important subject because there is a small handful of reports of 
little benefit arising from use of a 32-channel coil for fMRI, compared to a 
12-channel coil \citep{Hartwig, Kaza, Kaza2, Li}, and we have 
received anecdotal reports of poorer performance of the 32-channel coil for 
fMRI in our own center. Thus, we were motivated to investigate the RFC-MoCo 
effect because of a suspicion that it may produce a limiting motion sensitivity
when compared to the same protocol acquired with a smaller phased array coil.
Furthermore, recent reports \citep{Dijka, Power} have suggested 
that a significant (but unknown) fraction of 
correlations determined in resting-state functional connectivity (RS-FC) 
studies are driven by head motion. We hypothesized that the RFC-MoCo effect 
could be a cause of such errors, given the use of phased array coils for the 
majority of RS-FC fMRI studies to date.

The use of a prescan normalization as a means for correcting the RFC-MoCo 
effect has been studied by Hartwig et al. \citep{Hartwig}. Also Kaza et al. 
\citep{Kaza} used prescan-normalized and unnormalized data in a comparison 
of fMRI efficacy using 12-channel and 32-channel phased array coils. However, 
experimentally isolating the RFC-MoCo effect from the other potential causes 
of motion-related effects in time-series EPI would be very difficult if not 
impossible. Therefore in this paper we undertake simulations of the RFC-MoCo 
effect, which by design incorporates an assumption of perfect motion 
correction, that will allow us to elucidate the effect's impact upon
EPI time series data.

When choosing a receive coil for fMRI it is important to assure that the
subject within the coil will have visual access to stimuli or cues presented
on a screen placed outside the coil. Commercially available 12-channel arrays 
give good visual access for fMRI applications and are widely used. A birdcage 
coil with sixteen legs will give comparable visual access to that of the 
12-channel array consisting of the usual overlapping coil elements. This 
specification motivates our choice of coil geometries in our simulations. In 
this paper we carry out simulations of the RFC-MoCo effect for: (1) A 
12-channel head coil receive array consisting of 12 independent coil elements 
with a cylindrical geometry similar to those commonly used in modern scanners; 
and (2) A 16-leg receive-only birdcage head coil. In particular we will 
simulate: (1) The RFC-MoCo effect size; (2) A spatial map of temporal 
correlations due to the RFC-MoCo effect with respect to a seed point (as is 
often calculated in resting state fMRI analyses); and (3) The tSNR of an EPI 
time series when the noise of the RFC-MoCo effect is present in addition to 
the usual Gaussian noise of the complex-valued images.

\section{Theory}
\label{theory}

The use of multichannel receiver arrays is commonplace in magnetic resonance 
imaging and fMRI in particular. Whether the image data is to be used in 
accelerated GRAPPA-like parallel imaging or non-accelerated imaging, one must 
combine the image data from the individual coil elements into a single 
composite image. Of the various methods that could be used to generate a 
composite image from the images generated by each coil, the sum-of-squares (SOS)
method \citep{Roemer} is ubiquitous. The SOS image reconstruction method creates
a composite estimated image ${\hat \rho}({\bf r})$ from the complex-valued 
images $\rho_m({\bf r})$ associated with each of the $M$ receive coils of the 
array by the following operation:
\begin{eqnarray} 
  {\hat \rho}({\bf r}) =
  \left[
    \sum_{m=1}^M 
    \left|
      \rho_m({\bf r})
    \right|^2  
  \right]^{1/2}.
\label{eqn0}
\end{eqnarray} 
Note that when $M=1$ (a single channel) the result is the usual magnitude 
image associated with a birdcage coil.
Equation (\ref{eqn0}) assumes that we have a representation of the image 
${\hat \rho}({\bf r})$ in the continuum image-space rather than a discrete 
image-space. In practice the image will always be in a discrete image-space 
which is related to the underlying continuum image-space object through a
convolution with a point-spread function determined by the sampling of
$k$-space. The discretization is expected to influence the RFC-MoCo effect 
upon tSNR, as well as methods to correct for the effect, but for the purpose 
of establishing the size and importance of the effect the use of the continuum 
image-space representation should be sufficient and is expected to give a 
best-case estimate of the size and importance of the effect in EPI time 
series used in fMRI. 

In the absence of noise (electronic or body noise) and motion Equation 
(\ref{eqn0}) can be written as
\begin{eqnarray} 
  {\hat \rho}({\bf r}) =
  \left[
    \sum_{m=1}^M 
    \left|
      \rho({\bf r}) c_{m}({\bf r})
    \right|^2  
  \right]^{1/2}
  = |\rho({\bf r})|
  \left[
    \sum_{m=1}^M 
    \left|
      c_{m}({\bf r})
    \right|^2  
  \right]^{1/2}
  \label{eqn1}
\end{eqnarray} 
where $\rho({\bf r})$ is the complex valued true image and $c_m({\bf r})$ is 
the receive field of the $m^{th}$ coil element of the array. Therefore the SOS 
method results in an image contrast which is dependent upon the array
geometry and is spatially-fixed with respect to the scanner.

The imaged object $\rho$ may undergo rigid body motion (rotation and 
translation) with respect to the scanner frame of reference. We denote
this motion by a time-dependent affine transformation operator ${\mathcal A}
(t_n)$ where $t_n = n \Delta t$ is the time at which the $n^{th}$ volume of
image data is acquired. We will assume that the object does not move by a 
significant amount on a time scale less than $\Delta t$. This restriction 
imposes another best-case scenario on the effects of motion: Real motion is 
likely to occur within a TR period and will produce more complex effects, in 
particular spin history effects that will interact with the RFC-MoCo effect.
Thus, for simplicity, we consider the RFC-MoCo effect in isolation and note 
that the real effects on data are likely to be worse than presented here. The 
SOS estimated image of the object at time $t_n$ is then given by  
\begin{eqnarray} 
  {\hat \rho}({\bf r}, t_n) 
  &=&
  \left[
    \sum_{m=1}^M 
    \left|
      c_{m}({\bf r}) {\mathcal A} (t_n) \rho({\bf r})  + \eta_{m} 
    \right|^2  
  \right]^{1/2}
   \label{eqz1}
\end{eqnarray}
where $\eta_m$ is the uniform noise image for the $m^{th}$ coil. \footnote{If 
necessary the action of the affine transformation operator upon an image can 
be explicitly written as ${\mathcal A}\rho({\bf r}) = \rho({\bf R}{\bf r} - 
{\bf r}_o)$ where ${\bf R}$, a matrix representing rotation about the object's
center of mass, and ${\bf r}_o$, a vector representing translation of the 
center of mass, are functions of $t_n$.}

The usual, but idealized, motion correction is performed by applying the 
inverse of the affine transformation ${\mathcal A}(t_n)$ to the SOS image
${\hat \rho}({\bf r}, t_n)$ to obtain the motion-corrected image 
${\hat \rho}_c({\bf r}, t_n)$ given by
\begin{eqnarray} 
  {\hat \rho_c}({\bf r}, t_n)
  &=& {\mathcal A}^{-1} (t_n) {\hat \rho}({\bf r}, t_n)  \\
  &=& {\mathcal A}^{-1} (t_n)
  \left[
    \sum_{m=1}^M 
    \left|
      c_{m}({\bf r}) {\mathcal A} (t_n) \rho({\bf r})  + \eta_{m} 
    \right|^2  
  \right]^{1/2} \\
  &=& 
  \left[
    \sum_{m=1}^M 
    \left|
      \rho({\bf r}) {\mathcal A}^{-1}(t_n)  c_{m}({\bf r}) + \eta_{m} 
    \right|^2  
  \right]^{1/2}.
  \label{eqz1b}
\end{eqnarray}
Therefore the effect of combined motion and motion correction is mathematically
equivalent to moving the receive coil elements relative to the imaged object.
As explained in the Methods section this equivalence will be used to generate
the time varying contrast of our simulated RFC-MoCo effect. \footnote{If need 
be, the inverse affine transformation can be written more explicitly as 
${\mathcal A}^{-1} \rho({\bf r}) = \rho({\bf R}^{-1} ({\bf r}-{\bf r}_o))$.} 

To calculate the RFC-MoCo effect size, temporal correlation map, and tSNR we 
respectively calculate the percent difference map $D_p({\bf r})$, correlation 
value map $\chi({\bf r}_p,{\bf r}_{p'})$ and tSNR map ${\rm tSNR}({\bf r})$
defined according to:

\begin{eqnarray} 
  D_p({\bf r}) 
  = \frac{|{\hat \rho}_c({\bf r}, t_2) - {\hat \rho}_c({\bf r}, t_1)|}
  {{\hat \rho}_c({\bf r}, t_1)} \times 100
  \label{p_diff_1}
\end{eqnarray}

\begin{eqnarray} 
  \chi({\bf r}_p,{\bf r}_{p'}) = 
  \frac{{E[({\hat \rho}_c({\bf r}_{p}, t_n)-\mu_{{\bf r}_p})
           ({\hat \rho}_c({\bf r}_{p'}, t_n)-\mu_{{\bf r}_{p'}})]}} 
       {{\sigma_{{\bf r}_p}} \sigma_{{\bf r}_{p'}}} 
  \label{eq_corr}
\end{eqnarray}

\begin{eqnarray} 
  {\rm tSNR}({\bf r}) 
  &=& \frac{{\rm E}[{\hat \rho}_c({\bf r}, t_n)]}
  {\sqrt{{\rm E}[|{\hat \rho}_c({\bf r}, t_n)|^2]
  -|{\rm E}[{\hat \rho}_c({\bf r}, t_n)]|^2}} 
  \label{eqd}
\end{eqnarray}
where ${\rm E}$ denotes the expectation value with respect to time $t_n$ while
$\mu_{\bf r}$ and $\sigma_{\bf r}$ are respectively the mean and standard 
deviations of the simulated ${\hat \rho}_c({\bf r}, t_n)$ . 

When calculating the percent difference and the correlation map for our 
simulations we will neglect the noise term $\eta_{m}$. This results in the 
following quantities which are independent of the imaged object $\rho({\bf r})$:
\begin{eqnarray} 
  D_p({\bf r}) 
  = \frac{|C({\bf r}, t_2) - C({\bf r}, t_1)|}
  {C({\bf r}, t_1)} \times 100
  \label{p_diff_1a}
\end{eqnarray}
\begin{eqnarray} 
  \chi({\bf r}_p,{\bf r}_{p'}) = 
  \frac{{E[(C({\bf r}_{p}, t_n)-\mu_{{\bf r}_p})
           (C({\bf r}_{p'}, t_n)-\mu_{{\bf r}_{p'}})]}} 
       {\sqrt{E[(C({\bf r}_{p}, t_n)-\mu_{{\bf r}_p})^2]}
        \sqrt{E[(C({\bf r}_{p'}, t_n)-\mu_{{\bf r}_p'})^2]}} 
  \label{eq_corr_100}
\end{eqnarray}
where the net receive field contrast $C({\bf r}, t_n)$ is given by
\begin{eqnarray}
  C({\bf r}, t_n) 
  &=& 
  \left(
    \displaystyle \sum_{m=1}^M 
    \left|
      {\mathcal A}^{-1}(t_n)  c_{m}({\bf r})
    \right|^2  
  \right)^{1/2} .
  \label{eq_corr_101}
\end{eqnarray}
When calculating the simulated tSNR we will assume that the image is uniform,
$\rho({\bf r})= 1$, so that we may isolate the RFC-MoCo effect from brain 
contrast effects. In this case we may then write
\begin{eqnarray}
  \hspace{-27pt}
  {\rm tSNR}({\bf r})  &=&  \nonumber \\
  & & \hspace{-70pt} \frac{
    {\rm E} 
    \left[
      \left(
        \displaystyle \sum_{m=1}^M 
        \left|
          {\mathcal A}^{-1}(t_n)  c_{m}({\bf r}) + \eta_{m} 
        \right|^2  
      \right)^{1/2}
    \right]
  }{
    \sqrt{{\rm E}
    \left[
      \displaystyle \sum_{m=1}^M 
      \left|
        {\mathcal A}^{-1}(t_n)  c_{m}({\bf r}) + \eta_{m} 
      \right|^2  
    \right]
    -
    \left| {\rm E} 
      \left[ 
        \sqrt{
          \displaystyle \sum_{m=1}^M 
          \left|
            {\mathcal A}^{-1}(t_n)  c_{m}({\bf r}) + \eta_{m} 
          \right|^2
        } 
      \right]
    \right|^2}
  } .
  \label{eqd2}
\end{eqnarray}

When the motion of the object is a small translation, that is ${\bf r}_o(t) =
\delta {\bf r}(t)$ (see \ref{pert}) we can approximate Equations 
(\ref{p_diff_1a}) and (\ref{eq_corr_100}) by
\begin{eqnarray} 
  D_p({\bf r}) 
  = \frac{\left|[\delta {\bf r}(t_2) - \delta {\bf r}(t_1)] \cdot 
          \nabla C_{sos}({\bf r}, t_2) \right|}
  {C_{sos}({\bf r}, t_1)} \times 100
  \label{p_diff_2}
\end{eqnarray}
\begin{eqnarray} 
  \chi({\bf r}_p,{\bf r}_{p'}) = 
  \frac{{E[(\delta {\bf r}(t_n) \cdot \nabla C_{sos}({\bf r}_{p}, 0))
     (\delta {\bf r}(t_n) \cdot \nabla C_{sos}({\bf r}_{p'}, 0) )]}} 
     {\sqrt{E[(\delta {\bf r}(t_n) \cdot \nabla C_{sos}({\bf r}_{p}, 0))^2]}
     \sqrt{E[(\delta {\bf r}(t_n) \cdot \nabla C_{sos}({\bf r}_{p'}, 0))^2]}},
  \label{eq_corr_103}
\end{eqnarray}
where in deriving Equation (\ref{eq_corr_103}) we have assumed that 
$E[\delta {\bf r}(t_n)]=0$ as will be the case in our simulations. 

Although Equations (\ref{p_diff_2}) and (\ref{eq_corr_103}) will not be used in
our simulations of the RFC-MoCo effect these equations can be helpful with 
respect to understanding the effect. For example Equation (\ref{p_diff_2}) 
shows that the percent difference should be directly proportional to the  
translational displacement and the gradient of the SOS contrast field. Also 
from Equation (\ref{eq_corr_103}) it should be clear that if the small 
translations are in only one direction then the correlation map takes the 
value $1$ or $-1$ depending upon the sign of $\nabla C_{sos}({\bf r}_{p})$ and 
$\nabla C_{sos}({\bf r}_{p'})$ at the seed and map points respectively. When 
the motion in each dimension is temporally different then $-1 \le 
\chi({\bf r}_p,{\bf r}_{p'}) \le 1$ and the correlation map can have a more 
interesting structure, as will be seen in the results section of this paper.

\section{Methods}
\label{methods}

In this section we: (1) Describe the calculation of the simulated temporally 
varying receive field contrast for the 16-leg birdcage and 12-channel array 
head coils; (2) Describe the calculation of the metrics of the RFC-MoCo effect 
- percent difference, temporal correlation and tSNR maps - in a representative
transverse (orthogonal to the long axis of the head coil) plane at $z=90$ mm. 
Each coil has a radius of 130 mm at the plane of interest and the metrics 
are calculated over a 110 mm radius region centered in this plane. This is 
equivalent to assuming a simulated spherical phantom having a diameter 
of 220 mm - comparable to a typical image field-of-view for an adult human 
head - which is centered within the head coil. A uniform signal intensity from 
all points in space was assumed, thereby allowing us to investigate the 
RFC-MoCo effect in the absence of image contrast, transmission field 
heterogeneity and magnetic susceptibility gradient effects.

\subsection{Simulating the Temporally Varying RFC}

To calculate the receive fields of the head coils we construct the coil or coil
elements as a composite of line segments of current. As shown in  
\ref{B_line_seg} the receive field due to a single line segment with specified 
endpoints and current amplitude can be written in an exact mathematical form 
by means of a Biot-Savart Law integration. Then to translate the receive field 
at time $t_n$, thereby creating a time series of simulated RFC-MoCo data, we 
simply translate each endpoint of the line segments comprising the head coil or coil elements. All calculations were performed using C++ code and since the 
mathematical form of the receive field is known exactly then all calculations 
will be exact to within machine precision. Note that for both head coils we 
will neglect the effects of coil loading and mutual inductance between 
current-carrying line segments. Although the error due to these approximations 
is significant in the precise design and operation of receive coils it should 
not change the results given in this paper significantly.

\begin{figure}[htbp]
 \centering
 \includegraphics[width=380pt, height=220pt, viewport = 5 5 380 255]{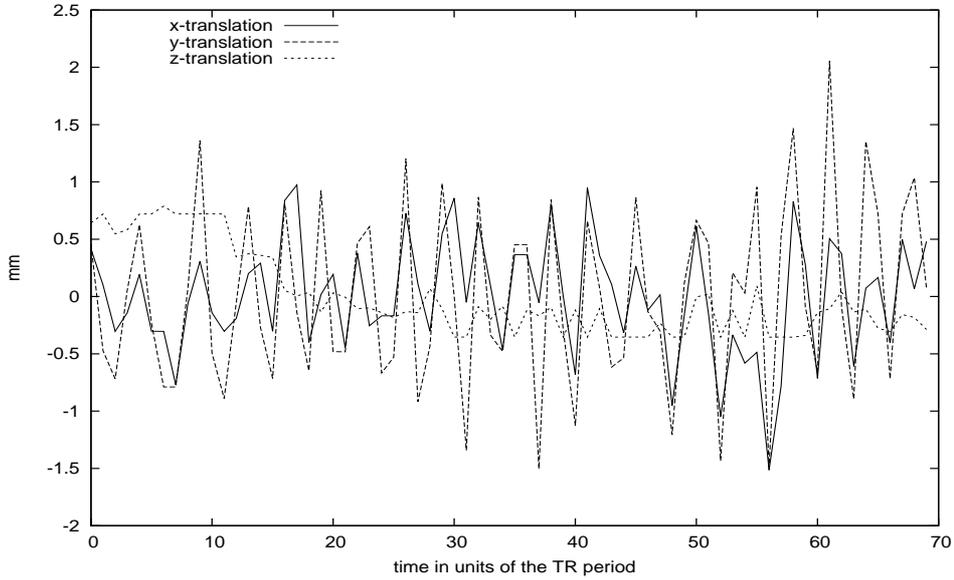}
 \caption{The time series of translational displacements used in the tSNR and
          correlation value calculations.}
 \label{displacement_time_series}
\end{figure}

Some of the simulations will be performed using a time series of realistic 
translational head motion data. Figure \ref{displacement_time_series} shows 
this realistic center-of-mass translation as a function of time. This motion 
data was obtained from the output of the FSL (FMRIB Software Library, 
University of Oxford, UK.) motion correction algorithm (MCFLIRT) \citep{MCFLIRT}
for a human subject in an actual fMRI experiment and is representative of 
motion data obtained on normal adult subjects at 3 T. The temporal mean of the 
motion was subtracted from the data in accordance with the assumptions of our 
motion model. To test the effects of variable overall motion the motion data 
is variably scaled to yield specific amounts of root-mean-square motion.

\vspace{8pt}
\noindent {\bf 16-Leg Birdcage Coil}
\vspace{4pt}

\begin{figure}[htbp]
 \centering
 \includegraphics[scale=0.35,viewport = 150 -10 300 565]{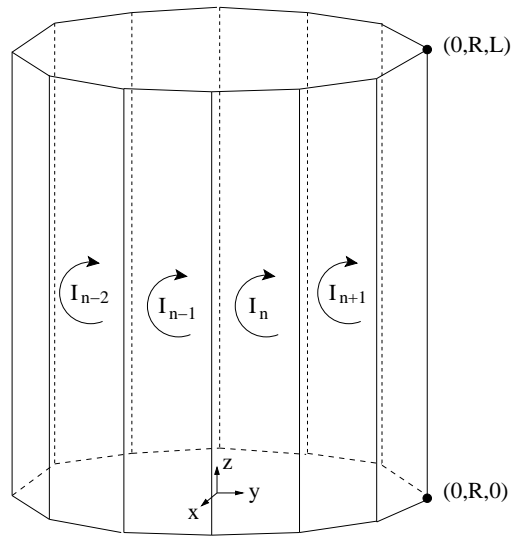}
 \caption{The simulations are done with a 16-leg birdcage coil but here we 
         depict a 12-leg coil to avoid clutter in the figure and clearly 
         convey the general coil geometry. The coil is depicted with mesh 
         currents $I_n$ $(n=1, \ldots, N)$ where $N$ is the number of legs.
         Note that the lower end ring lies in the $z=0$ mm plane.}
 \label{birdcage_coil}
\end{figure}

Figure \ref{birdcage_coil} depicts a birdcage coil of the type used in this
paper. In mode 1, the most spatially homogeneous mode \citep{Jin} the 
time-independent part of the mesh current $I_n$ is given by:
\begin{eqnarray} 
  I_n = C \cos(2\pi n/N)
  \label{bird_1}
\end{eqnarray}
where $N$ is the number of legs (struts) in the coil and C is a constant which
will not be of consequence in this paper. The current $\mathcal{I}_n$ in the 
$n^{th}$ leg is given by $\mathcal{I}_n = I_n - I_{n-1}$. 

\begin{figure}[htbp]
 \centering
 \includegraphics[scale=1.3, viewport = 50 20 300 220]{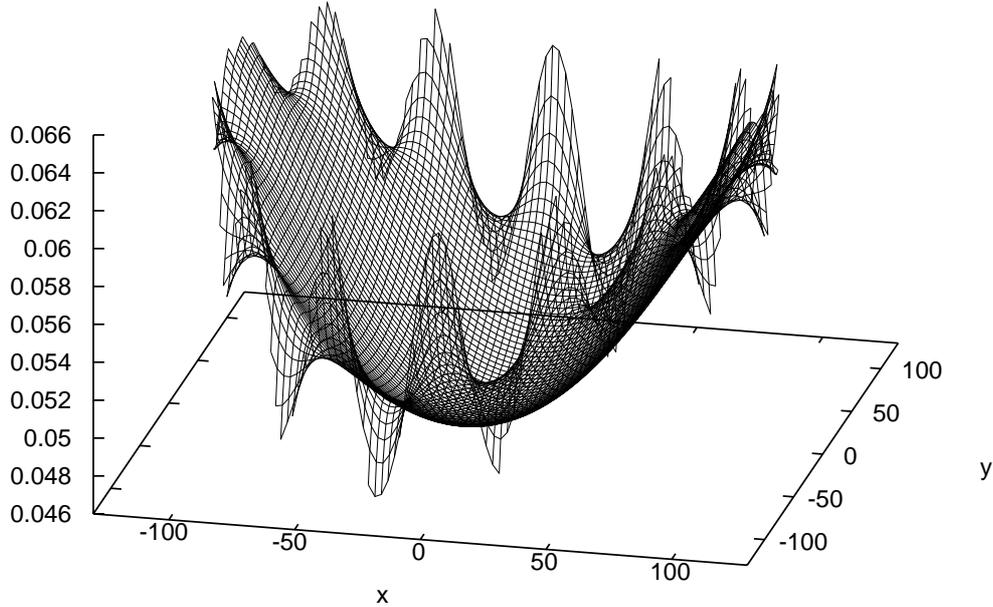}
 \caption{Receive field magnitude for 16-leg birdcage coil geometry (coil 
          radius = 130 mm, coil length $L = 186$ mm). The field is plotted 
          over a region of radius 110 mm centered in the transverse plane
          at $z=90$ within the coil. The receive field magnitude is plotted 
          as a unitless quantity since the homogeneity of the field is our 
          primary concern in this paper. }
 \label{birdcage_16_leg_rx_field}
\end{figure}

The endpoints of line segments comprising the legs and endrings of the birdcage
coil are arranged symmetrically on a cylindrical surface of 130 mm radius. 
Figure \ref{birdcage_16_leg_rx_field} shows the receive field for such a 
birdcage coil. The field is plotted over an axially-sliced region of a 
110 mm radius which is centered transversely within the coil. The important 
thing to note about Figure \ref{birdcage_16_leg_rx_field} is that the birdcage 
coil gives a receive field which varies by approximately $30\%$ over the 
region shown.

\vspace{8pt}
\noindent {\bf 12-Channel Array}
\vspace{4pt}

\begin{figure}[htbp]
 \centering
 \includegraphics[scale=0.8, viewport = 80 -5 380 250]{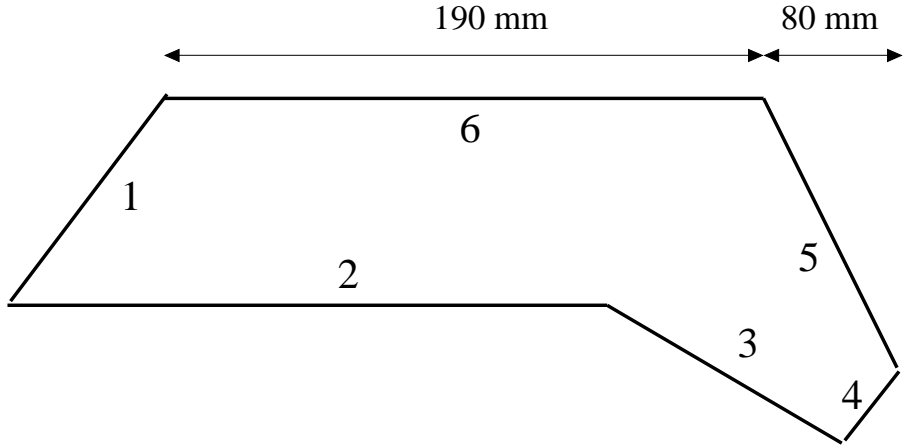}
 \caption{The general shape of a single coil element for the 12-element array 
          coil. The length of segments 2 and 6 are the same for all elements.
          Segments 1 lie in the $z=0$ transverse plane.}
 \label{coil_element_12_chan}
\end{figure}

\begin{figure}[htbp]
 \centering
 \includegraphics[scale=0.6, viewport = 0 0 380 260]{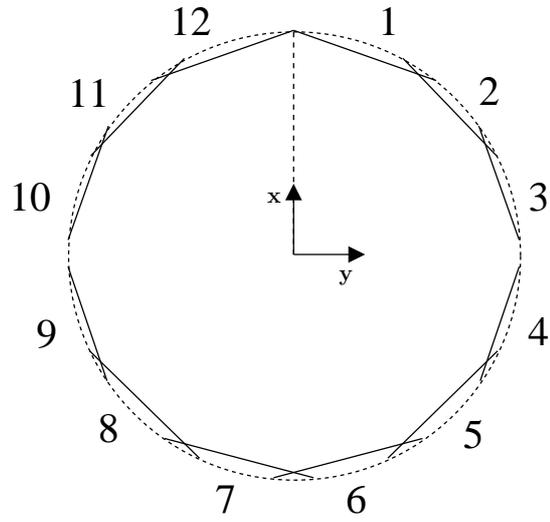}
 \caption{Cross-section of the simulated 12-channel receiver array showing
          length and orientations for segment 1 of each coil element. The 
          z-axis (parallel to $B_0$), of a right-handed coordinate system, 
          points into the plane of the page.}
 \label{12_chan}
\end{figure}

The simulated 12-channel array consists of $M=12$ coil elements of the general
form shown in Figure \ref{coil_element_12_chan} in which the coil element is
comprised of six segments each with a unit current in the clockwise direction
from the perspective of an observer external to the array. Figure \ref{12_chan}
depicts segment 1 of each coil element, which are taken to lie in a single 
transverse plane (the x,y-plane). The endpoints of the six line segments 
comprising each coil element lie on a cylindrical surface of 130 mm radius 
(see \ref{geom_12_chan}) and closely approximate the geometry of 
commercially available head coils. The sum of the fields due to each of the 
line segments comprising a given coil element yields the receive field of that 
coil element. The final SOS receive field is then calculated from the receive 
fields of the individual coil elements according to Equation (\ref{eqn1}).

\begin{figure}[htbp]
 \centering
 \includegraphics[scale=1.3, viewport = 70 20 280 200]{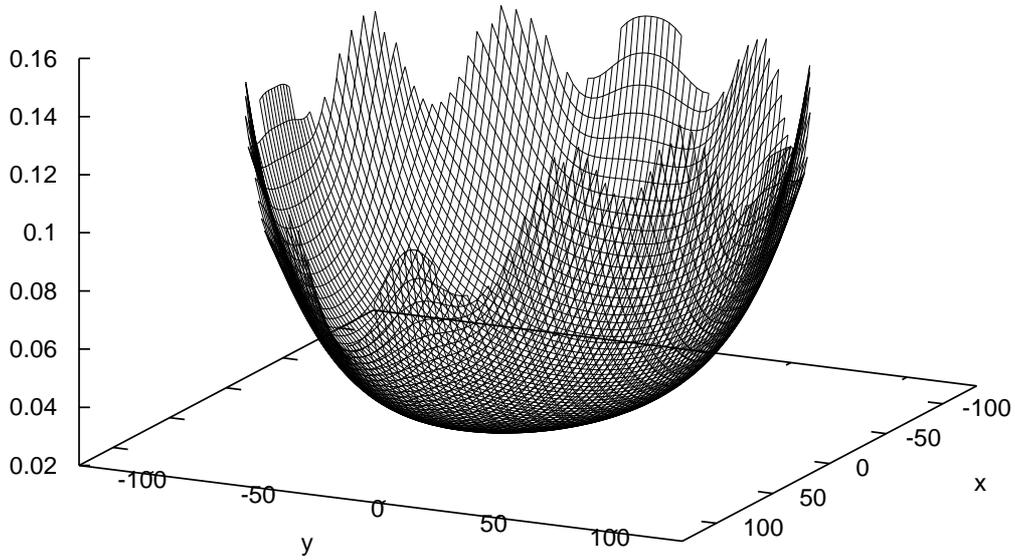}
 \caption{SOS receive field over an axially-sliced region of radius 110 mm 
          which is centered in the transverse plane at $z=90$ within the 
          12-element receive array. The receive field magnitude is 
          plotted as a unitless quantity since the homogeneity of the field is 
          our primary concern in this paper.}
 \label{12_chan_sos}
\end{figure}

Figure \ref{12_chan_sos} shows the SOS receive field for the simulated 
12-channel array. The field is plotted over an axially-sliced region of 110 mm 
radius which is centered transversely within the receive array. The important 
thing to note about Figure \ref{12_chan_sos} is that the 12-channel array 
gives a receive field which varies by as much as $400\%$ over the region 
shown.

\subsection{Calculating RFC-MoCo Effect Metrics}

\vspace{8pt}
\noindent {\bf Percent Difference}
\vspace{4pt}

For both head coil geometries we estimate the magnitude of the RFC-MoCo effect 
size by calculating the percent difference map, given by Equation 
(\ref{p_diff_1a}), when the object is displaced translationally by $1$ mm in 
the y-direction. For both head coils we also calculate the average and maximum
of the percent difference map over the $110$ mm radius region when the object 
is displaced by $0$ to $3.5$ mm. For the percent difference calculations no
noise is added to the simulated data.

\vspace{8pt}
\noindent {\bf Temporal Correlation}
\vspace{4pt}

We examine potential effects upon temporal correlation maps due to the RFC-MoCo
effect for translational head motion in the 12-channel array only. These 
simulations are performed by using the time series of realistic head motion
variables (see Figure \ref{displacement_time_series}) and calculating temporal 
correlation maps according to Equation (\ref{eq_corr_100}). To calculate the 
temporal correlation maps we set the root-mean-square (rms) magnitude of the 
translational motion to 1 mm but it should be noted that for small translations
we expect, from Equation (\ref{eq_corr_103}), that the correlation maps will 
be independent of the magnitude of the motion. The correlation is between a 
seed point at $x=-104$ mm, $y=0$ mm and $z=90$ mm and all points within the 
same axially-sliced ($z=90$ mm) region of radius $110$ mm which is centered 
within the array. For the temporal correlation calculations no noise is added 
to the simulated data because our principal concern is elucidating the general 
features one might expect to see in a temporal correlation map when the 
RFC-MoCo effect is of significant size \footnote{The significance is 
established by the percent difference maps}.

\vspace{8pt}
\noindent {\bf tSNR}
\vspace{4pt}

For the 12-channel array only we investigate the spatial dependence of the 
tSNR maps for varying amounts of realistic translational head motion (see 
Figure \ref{displacement_time_series}) and spatially uniform Gaussian 
electronic noise $\eta_m$. The noise is applied to the complex-valued time 
series image data according to Equation (\ref{eqd2}) such that it yields a 
tSNR level, 80 at the center of the image, representative of the tSNR seen in 
a 3-minute $T2^*$-weighted EPI time series at 3 T. TSNR maps were 
obtained for rms motion values of $0.0$, $0.5$, $1.0$ and $2.5$ mm.

\section{Results}
\label{results}

\vspace{8pt}
\noindent {\bf Percent Difference}
\vspace{4pt}

\begin{figure}[htbp]
  \begin{overpic}[scale=1.3,viewport = 30 20 380 150]{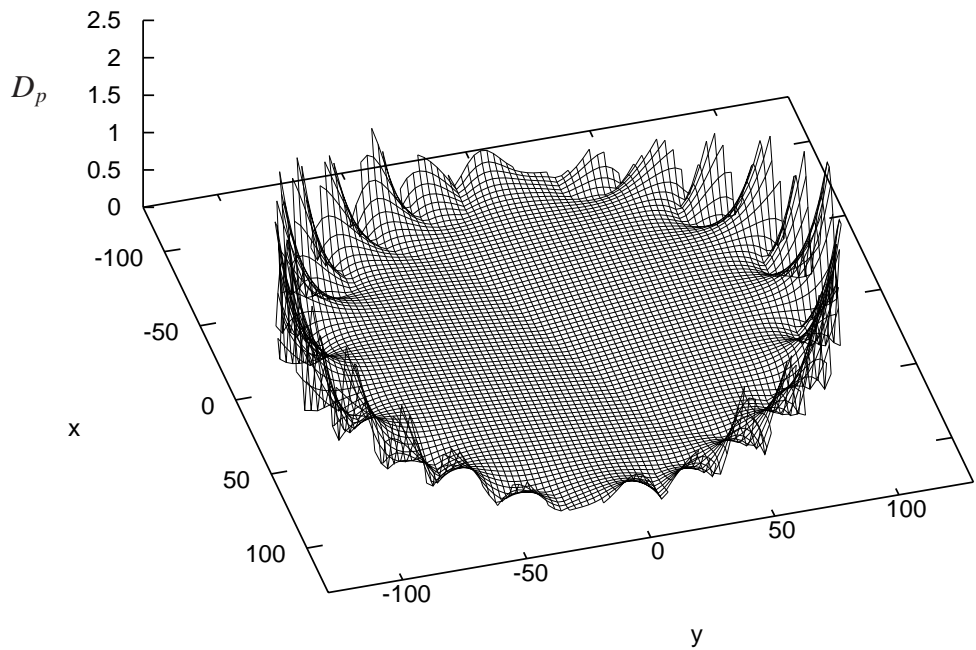}
    \put(-2,44){$D_p$}
  \end{overpic}
  \caption{Percent difference in receive field contrast due to 1 mm motion in 
          the $y$-direction for the 16-leg birdcage receive coil geometry 
          (radius = 130 mm). The percent difference is plotted over an 
          axially-sliced region of radius = 110 mm which is centered within the 
          array.}
  \label{birdcage_16_leg_percent_diff}
\end{figure}

\begin{figure}[htbp]
  \begin{overpic}[scale=1.3,viewport = 25 30 380 200]{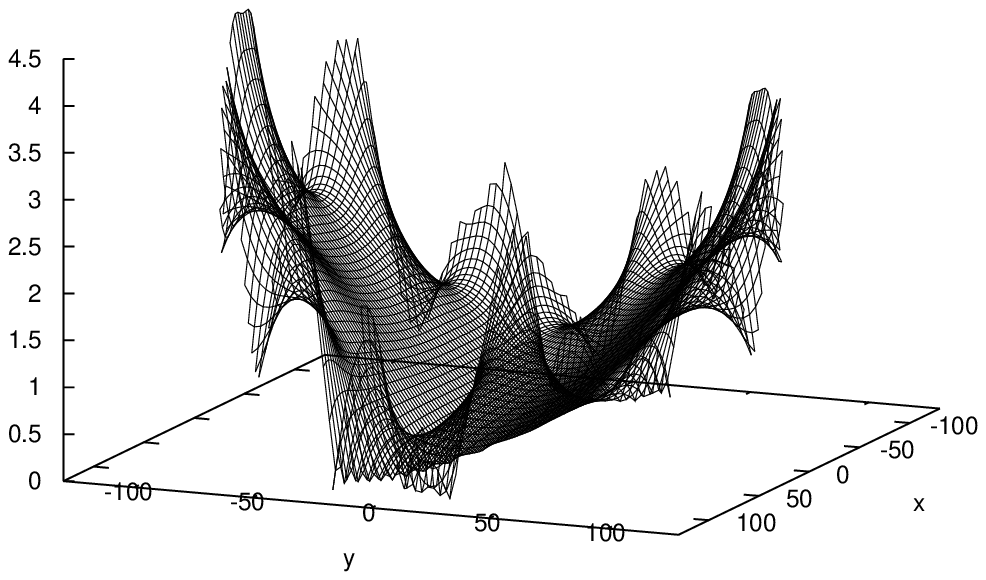}
    \put(-1,30){$D_p$}
  \end{overpic}
  \caption{Percent difference in receive field contrast due to 1 mm motion in 
          the $y$-direction for the 12-channel receiver array geometry (radius
          = 130 mm). The percent difference is plotted over an axially-sliced 
          region of radius = 110 mm which is centered within the array.}
 \label{12_chan_percent_diff}
\end{figure}

Figures \ref{birdcage_16_leg_percent_diff} and \ref{12_chan_percent_diff} show 
the percent difference in the receive field contrast due to a 1 mm translation
in the $y$-direction for the 16-leg birdcage and 12-channel array head coils,
respectively. From these figures it should be clear that this particular 
birdcage coil geometry yields significant reduction in the RFC-MoCo effect
as compared to the 12-channel array. It is also clear that the RFC-MoCo effect
for the 12-channel array is of similar magnitude if not greater than the 
percent difference expected from the BOLD effect at most points in this axial 
slice. Hence the effect should be an important systematic error in fMRI 
analysis if an array coil is used as the receiver. 

\begin{figure}[htbp]
 \centering
 \includegraphics[scale=1.05,viewport = 0 0 380 200]{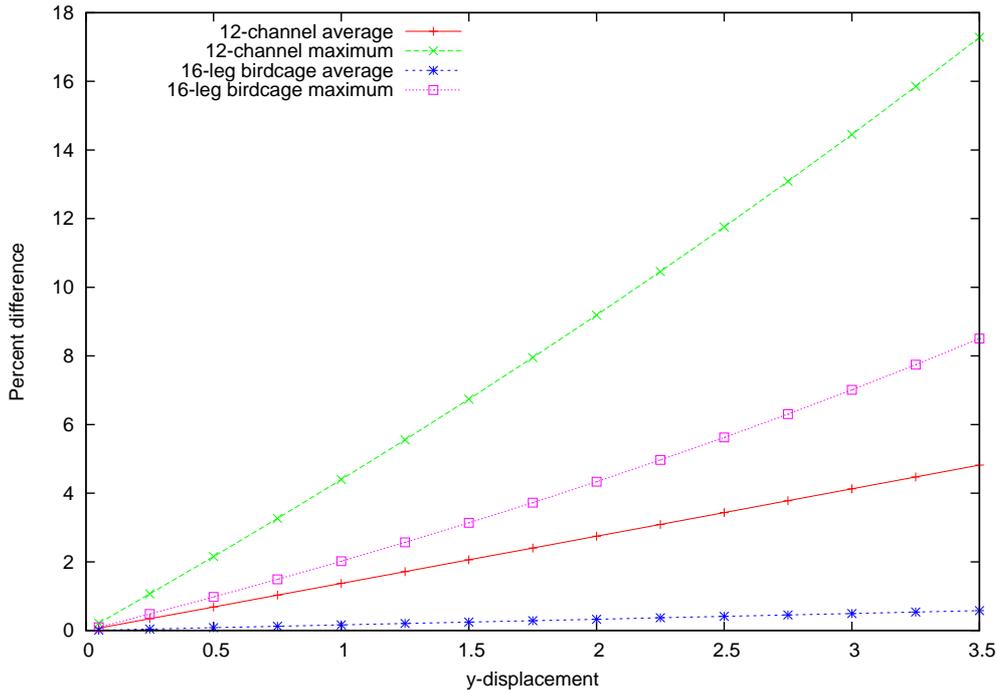}
 \caption{Comparison of percent difference between 16-leg birdcage and 
          12-channel array receiver head coils. Average and 
          maximum percent difference over an axially-sliced region of radius 
          = 110 mm which is centered within each coil are plotted versus
          displacement in the y-direction.}
 \label{percent_diff_vs_y_disp}
\end{figure}

Figure \ref{percent_diff_vs_y_disp} shows a plot of the average percent 
difference and the maximum percent difference over the $110$ mm radius region 
of an axial slice at $z=90$ mm versus the displacement in the y-direction. The 
trend is approximately linear, as would be expected for small displacements
(see Equation (\ref{eq_corr_102})).

\vspace{8pt}
\noindent {\bf Temporal Correlation}
\vspace{4pt}

\begin{figure}[htbp]
 \centering
 \begin{picture}(200,195)
 \put(110,260){\textcolor{red}{$\medbullet$}}
 \put(-85,220){\Large $\chi$}
 \includegraphics[scale=1.4,viewport = 100 20 380 180]{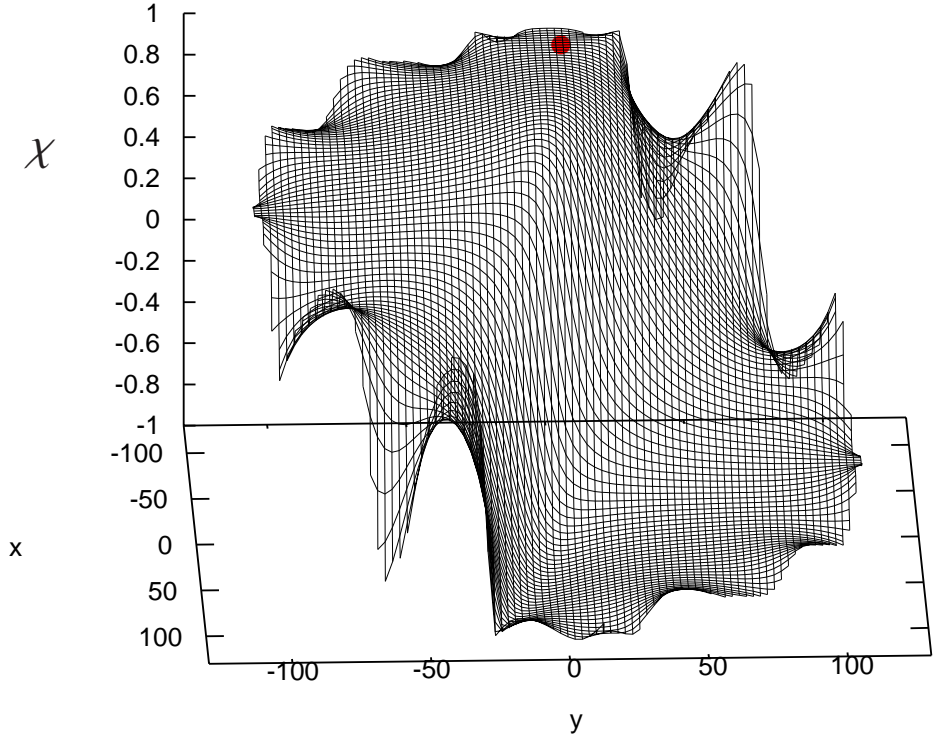}
 \end{picture}
 \caption{A surface plot showing the temporal correlation value 
          $\chi({\bf r}_p,{\bf r}_{p'})$ as a function position within an axial
          slice. The correlation is between a seed point ${\bf r}_{p'}$ at 
          $x=-104$ mm, $y=0$ mm and $z=90$ mm and all points ${\bf r}_p$ within 
          the same axially-sliced ($z=90$ mm) region of radius = 110 mm which 
          is centered within the array. For reference the value of $\chi$ at 
          the seed point is indicated by a red dot. The motion was purely 
          translational (1 mm rms) and taken from the ouput of the FSL (FMRIB's
          Software Library) MCFLIRT motion correction code (see Figure 
          \ref{displacement_time_series}).}
 \label{correlation_1}
\end{figure}

Figure \ref{correlation_1} is a surface plot showing the temporal correlation 
value as a function of location within an axial slice. The correlation is 
between a seed point at $x=-104$ mm, $y=0$ mm and $z=90$ mm and all points 
within the same axially-sliced ($z=90$ mm) region of radius $110$ mm which 
is centered within the array. For reference the value of $\chi$ at the seed 
point (1.0 as expected) is indicated by a red dot on the 3D surface. 
Notice the negative as well as positive correlation values.

\vspace{8pt}
\noindent {\bf tSNR}
\vspace{4pt}

Figures {\ref{tsnr_no_motion} through {\ref{tsnr_2.5mm}} show the tSNR surface
plots for the cases of no motion, $0.5$ mm, $1.0$ mm and $2.5$ mm rms motion 
for the 12-channel receive array. Color and viewpoint is different for each 
surface plot to aid in the visualization of the surface. Please refer to the 
axes for quantitative information. At 0.5 mm rms motion (Figure \ref{tsnr_0.5mm}) much of the tSNR benefit from the multiple channels is eliminated through 
the RFC-MoCo effect. At 1.0 mm rms motion (Figure \ref{tsnr_1.0mm}) the 
flattening of the tSNR map is effectively complete. At 2.5 mm rms motion 
(Figure \ref{tsnr_2.5mm}), which approximately corresponds to a pixel shift 
for the chosen electronic noise tSNR, the tSNR in regions nearest the coils is 
much reduced compared to that at the center of the array. Degradation of tSNR 
would be expected for all brain regions.

\begin{figure}[htbp]
 \centering
 \includegraphics[scale=1.3,viewport = 25 30 350 140]{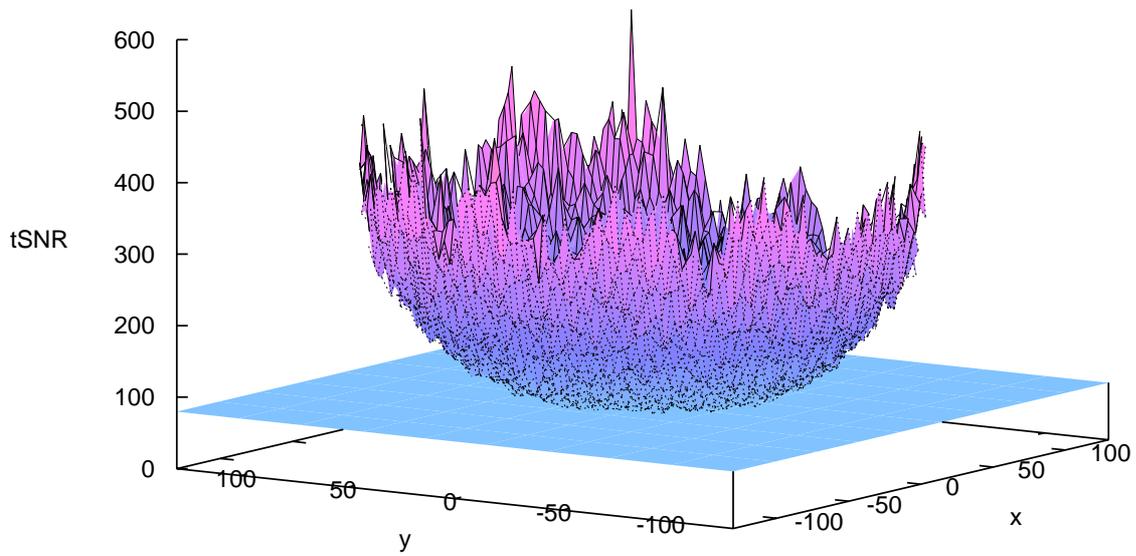}
 \caption{A surface plot showing the tSNR for the case of no motion as a 
          function of location within an axially-sliced ($z=90$ mm) region of 
          radius = 110 mm which is centered within the 12-element array. The 
          tSNR is 80 at the center of this region.}
 \label{tsnr_no_motion}
\end{figure}

\begin{figure}[htbp]
 \centering
 \includegraphics[scale=1.3,viewport = 25 35 350 185]{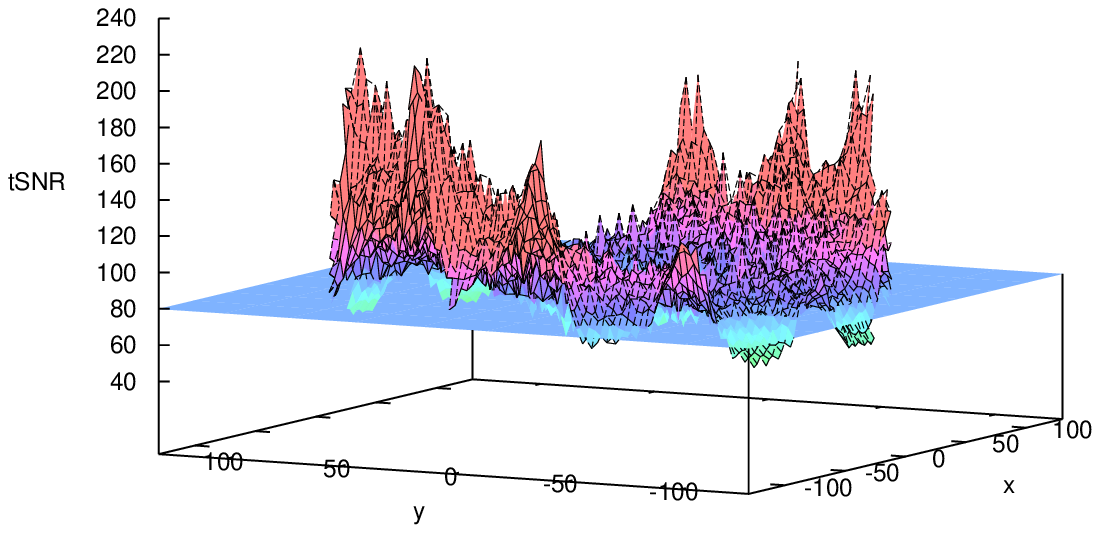}
 \caption{A surface plot showing the tSNR for the case of 0.5 mm rms motion 
          as a function of location within an axially-sliced ($z=90$ mm) 
          region of radius = 110 mm which is centered within the 12-element
          array. The motion was purely translational and taken from the ouput 
          of the FSL (FMRIB's Software Library) MCFLIRT motion correction code.}
 \label{tsnr_0.5mm}
\end{figure}

\begin{figure}[htbp]
 \centering
 \includegraphics[scale=1.3,viewport = 25 35 350 135]{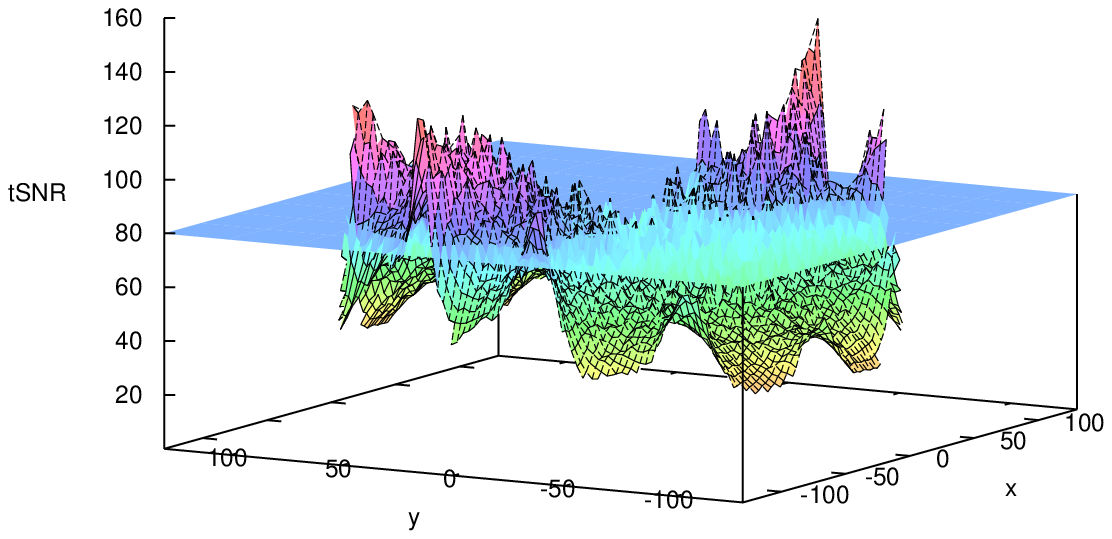}
 \caption{A surface plot showing the tSNR for the case of 1.0 mm rms motion as 
          a function of location within an axially-sliced ($z=90$ mm) region 
          of radius = 110 mm which is centered within the 12-element array. 
          The motion was purely translational and taken from the ouput of the 
          FSL (FMRIB's Software Library) MCFLIRT motion correction code.}
 \label{tsnr_1.0mm}
\end{figure}

\begin{figure}[htbp]
 \centering
 \includegraphics[scale=1.3,viewport = 25 35 350 200]{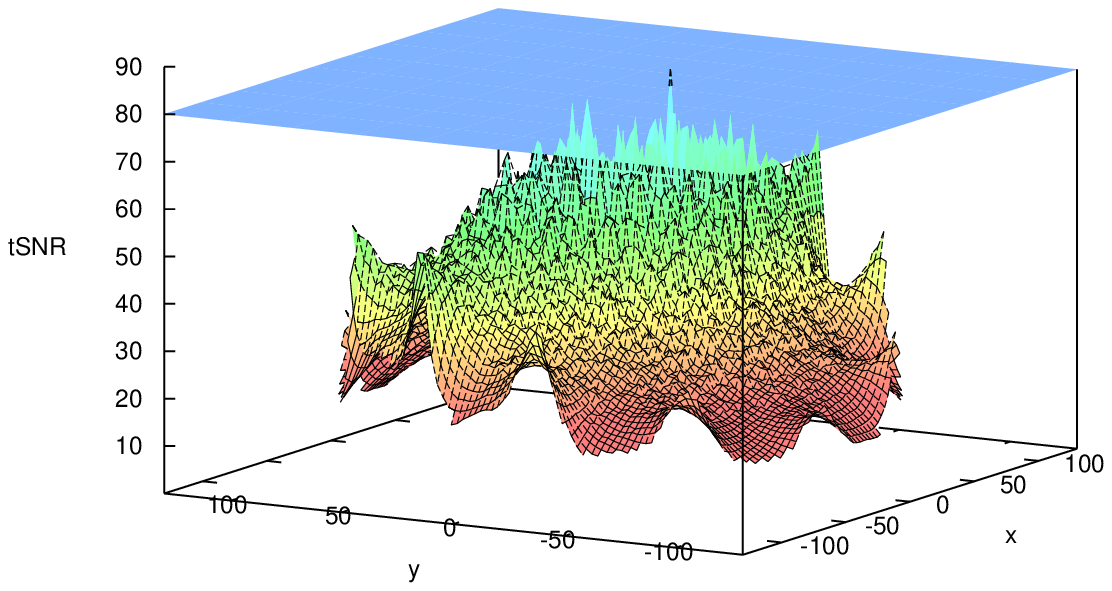}
 \caption{A surface plot showing the tSNR for the case of 2.5 mm (the size of
          a pixel typically associated with the tSNR at the center of the
          array) rms motion as a function of location within an axially-sliced 
          ($z=90$ mm) region of radius = 110 mm which is centered within the 
          12-element array. The motion was purely translational and taken from 
          the ouput of the FSL (FMRIB's Software Library) MCFLIRT motion 
          correction code.}
 \label{tsnr_2.5mm}
\end{figure}

\section{Discussion}
\label{discuss}

We set out to assess how receive field contrast could possibly limit the 
fidelity of motion-corrected time series MRI measurements acquired in the 
presence of significant subject motion. Simulations were performed so that the
specific interaction of the receive field contrast and motion could be 
assessed without the complicating factors, such as head-fixed image contrast 
and imperfect motion correction, that would likely arise in experimental data. 

We have shown in this work that: (1) The RFC-MoCo effect should be expected
to compete with the BOLD effect for a typical 12-channel cylindrical array,
(2) Interesting temporal correlations occur as a result of the RFC-MoCo effect
that will compete with the temporal correlations due to the BOLD effect and
(3) Moderate amounts of motion lead to serious tSNR degradation in regions 
of the 12-channel array receive field where, from consideration of the signal 
magnitude alone, one might expect improved tSNR. 

Caution should be exercised in interpreting the results given here for several
reasons:

\begin{enumerate}
\item{The effects of mutual inductance and coil loading were both neglected. 
We do not expect the neglect of these two complicating effects to do anything
but create greater local variability of the receive array SOS pattern and
hence even greater RFC-MoCo effect.}

\item{We have assumed the current-carrying conductors of the simulated head 
coils to be thin wires. The conductors of real head coils are thin tracings 
with a width (parallel to the cylidrical surface of the head coil) in the 
neighborhood of 3 mm. The width of the real conductors should have a 
spatially smoothing effect which would be expected to reduce the RFC-MoCo 
effect to some extent.}

\item{We used a root-sum-of-squares reconstruction of the final image. While
this method of reconstructing images from coil arrays is ubiquitous, other 
image combination methods are available such as the adaptive combination of 
coil images \citep{Ad_comb}, and these methods should perform differently with 
respect to the specific results of RFC-MoCo.}

\item{We have not simulated rotational motion of the head in this work and 
the presence of rotation should be expected to give even more interesting 
correlation maps.}

\item{Some manufactures combine the receive fields from the 12 coil elements 
in various manners to produce arrays that may be effectively smaller (eg. the 
"CP" and "Dual" receive modes of Siemens 3T Trio scanners) with correspondingly
more homogeneous receive fields. These coil combination approaches should 
establish reduced RFC-MoCo effects compared to the full multi-element array 
operation, but the remaining RFC-MoCo effect is likely to be larger than for a 
birdcage coil having similar physical dimensions.} 

\end{enumerate}

In a real fMRI experiment our assumption of perfect motion correction will not 
be satisfied and the degree to which the RFC-MoCo effect will influence the 
time series data will depend upon the accuracy of the applied motion correction
algorithm. When imperfect motion correction is applied there will
be a mixing of the various contrasts present in the image - those fixed 
relative to the head and those fixed relative to the scanner. Furthermore,
the presence of receive field contrast, as with other contrasts that are
fixed relative to the scanner, is expected to degrade motion correction 
performance through an underestimate of the actual motion - an effect we 
term the "anchoring effect." We are investigating the magnitude of the 
RFC-anchoring effect in parallel work. (Other scanner-fixed contrast mechanisms 
will generate their own anchoring effects.)

The question naturally arises as to the practical relevance of the results 
presented here for the RFC-MoCo effect. In the first instance, the 12-channel 
array was simulated in 
a manner that may not be appropriate for some scanner vendors. For example, in 
the absence of acceleration the default coil reception mode on a Siemens 
scanner would be "CP mode" \citep{Reykowski} which is designed to provide
near optimum SNR in the center region of the image. Thus, we cannot yet 
give a clear ranking of the RFC-MoCo effect's contribution to spurious 
correlations due to all causes in resting-state fMRI studies. We will attempt
to clarify this ranking in future simulation work.

Data processing methods such as independent component analysis \citep{ICA}
may be able to discriminate between artifactual and neurally-driven 
correlations in some circumstances, but it has not yet been established that 
the motion metrics output from the commonly used affine correction algorithms 
will capture sufficiently all the correlations introduced by the RFC-MoCo 
effect.  Recent work by Power et al. \citep{Power} suggests that some spurious 
correlations can be disambiguated by removing networks correlated with motion 
parameters, but further work is needed to establish the efficacy of this
approach.

In follow up simulations we will provide a similar analysis to that
presented in this work for the CP, dual, triple modes of the 12-channel
head array. We will also present a similar analysis for a 32-channel array
which has smaller somewhat hexagonally shaped coil elements arranged on the 
surface of a roughly spherical shell. Compared to the results given in this 
work we would expect the RFC-MoCo effect to be larger for the 32-channel array 
and smaller for dual and CP modes of the 12-channel array. We will then be 
in a better position to make quantitative estimates of the severity of the 
RFC-MoCo issue for connectivity studies from resting state data, and to 
compare the relative performance of 32-channel and 12-channel array coils 
for time series EPI under motion limiting regimes.

\newpage

\appendix

\section{Magnetic Field Due To An Arbitrarily Oriented Current Carrying Line 
Segment}
\label{B_line_seg}

In this section we derive the receive field due to a single line element. This 
result will be used to calculate the receive field of a coil element that can
be constructed from a set of such line elements. Figure \ref{line_element} 
depicts one such line element.

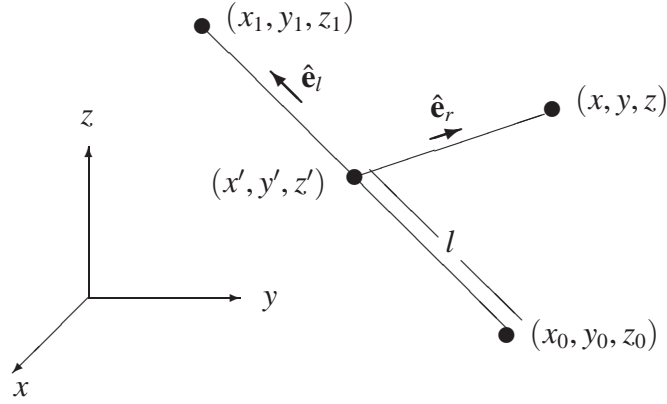
\begin{figure}
\setlength{\unitlength}{1mm}
\begin{picture}(93,46)
  \put(40,14){\vector(1,0){20}} 
  \put(63,13){$y$}
  \put(40,14){\vector(0,1){20}} 
  \put(39,37){$z$}
  \put(40,14){\vector(-1,-1){10}} 
  \put(30,1){$x$}
  \put(55, 50){\line(1, -1){40}}
  \put(75, 30){\line(3,1){25}}
  \put(104,39){$(x,y,z)$}
  \put(101,39){\circle*{2}}
  \put(56,28){$(x',y',z')$}
  \put(75,30){\circle*{2}}
  \put(98,8){$(x_0,y_0,z_0)$}
  \put(95,9){\circle*{2}}
  \put(58,50){$(x_1,y_1,z_1)$}
  \put(55,50){\circle*{2}}
  \put(86,22){\line(-1,1){9}} 
  \put(89,19){\line(1,-1){8}} 
  \put(87,20){$l$}
  \put(68,42){${\hat{\bf e}}_l$}
  \put(85,38){${\hat{\bf e}}_r$}
  \thicklines
  \put(68,40){\vector(-1,1){4}} 
  \put(85,35){\vector(3,1){4}} 
\end{picture}
\caption{A line element specified by the endpoints $(x_0,y_0,z_0)$ and 
         $(x_1,y_1,z_1)$ and carrying a unit current in the direction of the 
         unit vector ${\hat{\bf e}}_l$. The receive field is to be calculated 
         at points $(x,y,z)$.}
\label{line_element}
\end{figure}

The magnetic field ${\bf B}$ due to the line element can be calculated from the 
following form of the Biot-Savart Law:

\begin{eqnarray}
  {\bf B} = \int_0^L \frac{{\hat{\bf e}}_l \times {\hat{\bf e}}_r}{r^2} dl.
  \label{line_1}
\end{eqnarray}
The unit vectors ${\hat{\bf e}}_l$ and ${\hat{\bf e}}_r$ are given by
\begin{eqnarray}
  {\hat{\bf e}}_l 
   = \frac{\Delta x}{L} \; {\hat{\bf x}} + \frac{\Delta y}{L} \; {\hat{\bf y}} 
   + \frac{\Delta z}{L} \; {\hat{\bf z}} 
  \label{line_2}
\end{eqnarray}
where $\Delta x = x_1 - x_0$ etc and
\begin{eqnarray}
  {\hat{\bf e}}_r 
   = \frac{x - x'}{r} \; {\hat{\bf x}} + \frac{y - y'}{r} \; {\hat{\bf y}} 
   + \frac{z - z'}{r} \; {\hat{\bf z}} 
  \label{line_3}
\end{eqnarray}
where
\begin{eqnarray}
  L = 
   \sqrt{(\Delta x)^2 + (\Delta y)^2  + (\Delta z)^2}
  \label{line_4}
\end{eqnarray}
and
\begin{eqnarray}
  r = 
   \sqrt{(x - x')^2 + (y - y')^2  + (z - z')^2}.
  \label{line_5}
\end{eqnarray}
Since 
\begin{eqnarray}
  x' = x_0 + l {\hat{\bf e}}_l \cdot {\hat{\bf x}} = x_0 + (l/L) \Delta x \\
  y' = y_0 + l {\hat{\bf e}}_l \cdot {\hat{\bf y}} = y_0 + (l/L) \Delta y \\
  z' = z_0 + l {\hat{\bf e}}_l \cdot {\hat{\bf z}} = z_0 + (l/L) \Delta z 
  \label{line_6}
\end{eqnarray}
then
\begin{eqnarray}
  {\hat{\bf e}}_r 
   &=& r^{-1} [x - x_0 - (l/L) \Delta x] \; {\hat{\bf x}} \\
   &+& r^{-1} [y - y_0 - (l/L) \Delta y] \; {\hat{\bf y}} \\
   &+& r^{-1} [z - z_0 - (l/L) \Delta z] \; {\hat{\bf z}} 
  \label{line_7}
\end{eqnarray}
and
\begin{eqnarray}
  \hspace{-15pt}
  r = 
   \sqrt{[x - x_0 - l\Delta x /L]^2 + [y - y_0 - l\Delta y /L]^2  
   + [z - z_0 - l\Delta z /L]^2}.
  \label{line_8}
\end{eqnarray}
Since
\begin{eqnarray}
  \hspace{-25pt}
  {\hat{\bf e}}_l \times {\hat{\bf e}}_r \!\! &=&
  \left|
    \begin{matrix}
      {\hat{\bf x}} & {\hat{\bf y}} & {\hat{\bf z}} \\
      \frac{x_1 - x_0}{L} & \frac{y_1 - y_0}{L} & \frac{z_1 - z_0}{L}  \\
      \frac{x - x'}{r}  & \frac{y - y'}{r} &  \frac{z - z'}{r}    
    \end{matrix} 
  \right| \\
  &=& (Lr)^{-1} [\Delta y (z - z') - \Delta z (y - y')] \; {\hat{\bf x}} 
      \nonumber \\
  &+& (Lr)^{-1} [\Delta z (x - x') - \Delta x (z - z')] \; {\hat{\bf y}} 
      \nonumber \\
  &+& (Lr)^{-1} [\Delta x (y - y') - \Delta y (x - x')] \; {\hat{\bf z}} \\
  &=& (Lr)^{-1} [\Delta y (z - z_0 - l \Delta z /L)
                - \Delta z (y - y_0 - l \Delta y /L)] \; {\hat{\bf x}} 
      \nonumber \\
  &+& (Lr)^{-1} [\Delta z (x - x_0 - l \Delta x /L)
                - \Delta x (z - z_0 - l \Delta z /L)] \; {\hat{\bf y}} 
      \nonumber \\
  &+& (Lr)^{-1} [\Delta x (y - y_0 - l \Delta y /L)
                 - \Delta y (x - x_0 - l \Delta x /L)] \; {\hat{\bf z}} 
  \label{line_9}
\end{eqnarray}
then
\begin{eqnarray}
  B_x(x,y,z) 
  &=& \frac{\Delta y}{L} \int_0^L \frac{z - z_0 - l \Delta z/L}{r^3} dl
      - \frac{\Delta z}{L} \int_0^L \frac{y - y_0 - l \Delta y/L}{r^3} dl \\
  B_y(x,y,z) 
  &=& \frac{\Delta z}{L} \int_0^L \frac{x - x_0 - l \Delta x/L}{r^3} dl 
      - \frac{\Delta x}{L} \int_0^L \frac{z - z_0 - l \Delta z/L}{r^3} dl \\
  B_z(x,y,z) 
  &=& \frac{\Delta x}{L} \int_0^L \frac{y - y_0 - l \Delta y/L}{r^3} dl 
      - \frac{\Delta y}{L} \int_0^L \frac{x - x_0 - l \Delta x/L}{r^3} dl 
  \qquad \;\;
  \label{line_10}
\end{eqnarray}
where
\begin{eqnarray}
  r^2 &=& (x-x_0)^2 - 2(l/L)\Delta x (x-x_0) + (l/L)^2 (\Delta x)^2 
  \nonumber \\
  &+& (y-y_0)^2 - 2(l/L)\Delta y (y-y_0) + (l/L)^2 (\Delta y)^2
  \nonumber \\
  &+& (z-z_0)^2 - 2(l/L)\Delta z (z-z_0) + (l/L)^2 (\Delta z)^2
  \label{line_11} \\
  &=& (x-x_0)^2 + (y-y_0)^2 + (z-z_0)^2 \nonumber \\
  &-& 2(l/L)[(x-x_0) \Delta x + (y-y_0)\Delta y + (z-z_0)\Delta z] + l^2.
  \label{line_12}
\end{eqnarray}
Substituting
\begin{eqnarray}
  B_x(x,y,z) 
  &=& \frac{\Delta y}{L} 
  \int_0^L \frac{z - z_0 - l \Delta z/L}{[l^2 - \beta l + \kappa]^{3/2}} dl
  - \frac{\Delta z}{L} 
  \int_0^L \frac{y - y_0 - l \Delta y/L}{[l^2 - \beta l + \kappa]^{3/2}} dl \\
  B_y(x,y,z) 
  &=& \frac{\Delta z}{L} 
  \int_0^L \frac{x - x_0 - l \Delta x/L}{[l^2 - \beta l + \kappa]^{3/2}} dl 
  - \frac{\Delta x}{L} 
  \int_0^L \frac{z - z_0 - l \Delta z/L}{[l^2 - \beta l + \kappa]^{3/2}} dl \\
  B_z(x,y,z) 
  &=& \frac{\Delta x}{L} 
  \int_0^L \frac{y - y_0 - l \Delta y/L}{[l^2 - \beta l + \kappa]^{3/2}} dl 
  - \frac{\Delta y}{L} 
  \int_0^L \frac{x - x_0 - l \Delta x/L}{[l^2 - \beta l + \kappa]^{3/2}} dl 
  \qquad \;\;
  \label{line_13}
\end{eqnarray}
where
\begin{eqnarray}
  \kappa &=& (x-x_0)^2 + (y-y_0)^2 + (z-z_0)^2  \\
  \beta &=& (2/L)[(x-x_0) \Delta x + (y-y_0)\Delta y + (z-z_0)\Delta z].
  \label{line_14}
\end{eqnarray}
From the integral tables
\begin{eqnarray}
   \int \frac{d+ex}{[ax^2 + bx + c]^{3/2}} \; dx 
   = \frac{2bd - 4ce + (4ad - 2be)x}{(4ac - b^2)(ax^2 + bx + c)^{1/2}}
  \label{line_17}  
\end{eqnarray}
so that
\begin{eqnarray}
   \hspace{-40pt}
   \int_0^L \!\! \frac{d+ex}{[x^2 + bx + c]^{3/2}} \; dx 
   \! &=& \!\! \frac{2bd - 4ce + (4d - 2be)L}{(4c - b^2)(L^2 + bL + c)^{1/2}}
   + \frac{4ce - 2bd}{c^{1/2}(4c - b^2)}
  \label{line_17b} \nonumber \\  
  \! &=& \!\! \frac{(4L + 2b)d - (4c + 2bL)e}{(4c - b^2)(L^2 + bL + c)^{1/2}}
   + \frac{4ce - 2bd}{c^{1/2}(4c - b^2)}.
  \label{line_17c}  
\end{eqnarray}
We can then write
\begin{eqnarray}
  B_x 
  &=& \frac{\Delta y}{L} 
  \left[ 
    \frac{(4L-2\beta)(z-z_0) + (4 \kappa - 2 \beta L) \Delta z/L}
    {(4\kappa - \beta^2)(L^2 - \beta L + \kappa)^{1/2}}
    + \frac{2\beta(z-z_0) - 4\kappa \Delta z/L)}
           {\sqrt{\kappa}(4\kappa - \beta^2)}   
  \right] \nonumber \\
  &-& \frac{\Delta z}{L} 
  \left[ 
    \frac{(4L-2\beta)(y-y_0) + (4 \kappa - 2 \beta L) \Delta y/L}
    {(4\kappa - \beta^2)(L^2 - \beta L + \kappa)^{1/2}}
    + \frac{2\beta(y-y_0) - 4\kappa \Delta y/L)}
           {\sqrt{\kappa}(4\kappa - \beta^2)}   
  \right]  \nonumber \\
  &=& 
  \frac{(4L-2\beta)[\Delta y (z-z_0) - \Delta z(y-y_0)]}
  {L(4\kappa - \beta^2)(L^2 - \beta L + \kappa)^{1/2}} \nonumber \\
  &+& \frac{2\beta[\Delta y (z-z_0) - \Delta z(y-y_0)]}
         {L \sqrt{\kappa}(4\kappa - \beta^2)} 
  \label{line_18}
\end{eqnarray}
or
\begin{eqnarray}
  B_x &=& 
  \left[
    \frac{4L-2\beta}{(L^2 - \beta L + \kappa)^{1/2}}
    + \frac{2\beta}{\sqrt{\kappa}} 
  \right]
  \frac{\Delta y (z-z_0) - \Delta z(y-y_0)}{L(4\kappa - \beta^2)}.
  \label{line_18b}
\end{eqnarray}
Similarly
\begin{eqnarray}
  B_y &=& 
  \left[
    \frac{4L-2\beta}{(L^2 - \beta L + \kappa)^{1/2}}
    + \frac{2\beta}{\sqrt{\kappa}} 
  \right]
  \frac{\Delta z (x-x_0) - \Delta x(z-z_0)}{L(4\kappa - \beta^2)}
  \label{line_19}
\end{eqnarray}
and 
\begin{eqnarray}
  B_z &=& 
  \left[
    \frac{4L-2\beta}{(L^2 - \beta L + \kappa)^{1/2}}
    + \frac{2\beta}{\sqrt{\kappa}} 
  \right]
  \frac{\Delta x (y-y_0) - \Delta y(x-x_0)}{L(4\kappa - \beta^2)}. 
  \label{line_20}
\end{eqnarray}
Note, as a consistency check, that Equations (\ref{line_18}), (\ref{line_19}) 
and (\ref{line_20}) reduce to the field due to an infinite length wire with
unit current oriented along any of the axes \citep{arfken}. The receive field 
for a single line segment wire with unit current directed from point 
$(x_0,y_0,z_0)$ to point $(x_1,y_1,z_1)$ is then
\begin{eqnarray}
  B_r^* &=& B_x - i B_y \\
  &=& 
  \frac{(4L-2\beta)[\Delta y (z-z_0) - \Delta z(y-y_0)]}
  {L(4\kappa - \beta^2)(L^2 - \beta L + \kappa)^{1/2}} \nonumber \\
  &+& \frac{2\beta[\Delta y (z-z_0) - \Delta z(y-y_0)]}
         {L \sqrt{\kappa}(4\kappa - \beta^2)} \nonumber \\
  &-& 
  i \frac{(4L-2\beta)[\Delta z (x-x_0) - \Delta x(z-z_0)]}
  {L(4\kappa - \beta^2)(L^2 - \beta L + \kappa)^{1/2}} \nonumber \\
  &-& i \frac{2\beta[\Delta z (x-x_0) - \Delta x(z-z_0)]}
         {L \sqrt{\kappa}(4\kappa - \beta^2)}. 
  \qquad
  \label{line_21}
\end{eqnarray}

\section{12-Channel Array Geometry}
\label{geom_12_chan}

 The first end point $(x^0_{nm},y^0_{nm})$ and the second end
point $(x^1_{nm},y^1_{nm})$ of $m^{th}$ segment of the $n^{th}$ coil element 
are:

\vspace{8pt}
\noindent {\bf Top half of array}
\vspace{8pt}

\hspace{-25pt}
\begin{tabular}{l l}
   $x^0_{11} = r $ &  
   $y^0_{11} = 0$ \\
   $x^1_{11} = r \cos \phi_1$ & 
   $y^1_{11} = r \sin \phi_1$ \\
 $x^0_{21} = r \cos(\phi_1 - \theta_2)$ & 
   $y^0_{21} = r \sin(\phi_1 -\theta_2)$ \\
   $x^1_{21} = r \cos(\phi_2 + \phi_1 -\theta_2)$ & 
   $y^1_{21} = r \sin(\phi_2 + \phi_1 -\theta_2)$ \\
 $x^0_{31} = r \cos(\phi_2 + \phi_1 -\theta_2 - \theta_1) $ & 
   $y^0_{31} = r \sin(\phi_2 + \phi_1 -\theta_2 - \theta_1)$ \\
   $x^1_{31} = r \cos(\phi_3 + \phi_2 + \phi_1 -\theta_2 - \theta_1)$ & 
   $y^1_{31} =  r \sin(\phi_3 + \phi_2 + \phi_1 -\theta_2 - \theta_1)$ \\
 $x^0_{12,1} =  \cos \phi_1$ & 
   $y^0_{12,1} = - r \sin \phi_1$ \\
   $x^1_{12,1} = r$ &  
   $y^1_{12,1} = 0$ \\
 $x^0_{11,1} = r \cos(\phi_2 + \phi_1 -\theta_2)$ & 
   $y^0_{11,1} = -r \sin(\phi_2 + \phi_1 -\theta_2)$ \\
   $x^1_{11,1} = r \cos(\phi_1 - \theta_2)$ & 
   $y^1_{11,1} = -r \sin(\phi_1 - \theta_2)$ \\
 $x^0_{10,1} = r \cos(\phi_3 + \phi_2 + \phi_1 -\theta_2 - \theta_1)  $ & 
   $y^0_{10,1} = -r \sin(\phi_3 + \phi_2 + \phi_1 -\theta_2 - \theta_1) $ \\
   $x^1_{10,1} = r \cos(\phi_2 + \phi_1 -\theta_2 - \theta_1)$ & 
   $y^1_{10,1} = -r \sin(\phi_2 + \phi_1 -\theta_2 - \theta_1)$ \\
 \label{array_pts_1}
\end{tabular} 

\noindent where $\phi_n = 2 \sin^{-1}(l_n/2r)$ is the angle subtended by a 
chord of length $l_n$ giving the length of segment $1$ of the $n^{th}$ coil 
element and $\theta_n = 2 \sin^{-1}(d_n/2r)$ is the angle subtended by a chord
of length $d_n$ giving the chord length of the coil overlap and gaps. 

\vspace{8pt}
\noindent {\bf Bottom half of array}
\vspace{8pt}

\hspace{-25pt}
\begin{tabular}{l l}
 $x^0_{41} = - r \sin \theta_0$ &  
   $y^0_{41} = r \cos \theta_0$ \\
   $x^1_{41} = - r \sin (\phi_4 + \theta_0)$ & 
   $y^1_{41} = r \cos (\phi_4 + \theta_0)$ \\
 $x^0_{51} = -r \sin(\phi_4  + \theta_0 -\theta_1 )$ & 
   $y^0_{51} = r \cos(\phi_4  + \theta_0 -\theta_1)$ \\
   $x^1_{51} = -r \sin(\phi + \phi_4 + \theta_0 -\theta_1)$ & 
   $y^1_{51} = r \cos(\phi + \phi_4 + \theta_0 -\theta_1)$ \\
 $x^0_{61} = -r \sin(\phi + \phi_4 + \theta_0 -\theta_1 - \theta_2)$ & 
   $y^0_{61} = r \cos(\phi + \phi_4 + \theta_0 -\theta_1 - \theta_2)$ \\ 
   $x^1_{61} = -r \sin(2\phi + \phi_4 + \theta_0 -\theta_1 - \theta_2)$ & 
   $y^1_{61} = r \cos(2\phi + \phi_4 + \theta_0 -\theta_1 - \theta_2)$ \\
 $x^0_{71} = -r \sin(2\phi + \phi_4 + \theta_0 -\theta_1 - 2\theta_2)$ & 
   $y^0_{71} = r \cos(2\phi + \phi_4 + \theta_0 -\theta_1 - 2\theta_2)$ \\
   $x^1_{71} = -r \sin(3\phi + \phi_4 + \theta_0 -\theta_1 - 2\theta_2)$ &  
   $y^1_{71} = r \cos(3\phi + \phi_4 + \theta_0 -\theta_1 - 2\theta_2)$ \\
 $x^0_{81} = -r \sin(3\phi + \phi_4 + \theta_0 -\theta_1 - 3\theta_2)$ & 
   $y^0_{81} = r \cos(3\phi + \phi_4 + \theta_0 -\theta_1 - 3\theta_2)$ \\
   $x^1_{81} = -r \sin(4\phi + \phi_4 + \theta_0 -\theta_1 - 3\theta_2)$ & 
   $y^1_{81} = r \cos(4\phi + \phi_4 + \theta_0 -\theta_1 - 3\theta_2)$ \\
 $x^0_{91} =  -r \sin(4\phi + \phi_4 + \theta_0 - 2\theta_1 - 3\theta_2)$ & 
   $y^0_{91} = r \cos(4\phi + \phi_4 + \theta_0 - 2\theta_1 - 3\theta_2)$ \\ 
   $x^1_{91} = - r \sin \theta_0 $ & 
   $y^1_{91} = - r \cos \theta_0$ \\ 
 \label{array_pts_2}
\end{tabular}

\noindent where $\phi$, the angle subtended by segment 1 for coils 5 through 
8, is determined by the condition $\pi - 2(\phi_4 + \theta_0 - \theta_1) = 
4\phi - 3\theta_2$ or $\phi = \pi/4 - (\phi_4 + \theta_0 - \theta_1)/2 
+ 3\theta_2/4$. We choose $r = 130$, $d_0 = 7$, $d_1 = 20$, $d_2 = 23$, 
and $l_1 = 95$, $l_3 = l_4 = l_{10} = l_9 = 77$. For segment 4 we choose 
$r=50$ with all the same angles as for segment 1. See Figure 
\ref{coil_element_12_chan} for the $z$-coordinates of each endpoint.

\section{Small Rotations and Translations}
\label{pert}

Let $f({\bf r})$ be a scalar function of the vector ${\bf r}$ and let ${\bf r}' 
= {\bf R}^T({\bf r} - {\bf r}_o)$. For infinitesimal rotations and 
translations we can write ${\bf r}' = {\bf r} - \delta {\bf r} - \delta 
{\boldsymbol \phi} \times {\bf r}$ where $\delta {\bf r}$ is an infinitesimal
translation and $\delta {\boldsymbol \phi}$ is a vector giving the orientation 
of the axis of rotation and the magnitude of an infinitesimal angle of 
rotation about this axis. We want to approximate $f({\bf r}')$ in terms of the
$\nabla f$ and small translational and rotational variables. For ${\bf r}' = 
{\bf r} + {\bf d}$ we may always write:
\begin{eqnarray}
  f({\bf r}') \approx f({\bf r}) + {\bf d} \cdot \nabla f({\bf r})
  \label{chain_0}
\end{eqnarray}
hence
\begin{eqnarray}
  f({\bf r}') \approx f({\bf r}) - (\delta {\bf r} + \delta {\boldsymbol \phi} 
  \times {\bf r}) \cdot \nabla f({\bf r})
  \label{chain_1}
\end{eqnarray}
which can be written as
\begin{eqnarray}
  f({\bf r}') \approx f({\bf r}) - \delta {\bf r} \cdot \nabla f({\bf r})
  - \delta {\boldsymbol \phi} \cdot {\bf r} \times \nabla f({\bf r}).
  \label{chain_2}
\end{eqnarray}
Therefore for small translations only we may write the time-varying net
receive field as
\begin{eqnarray} 
  C_{sos}({\bf r}_{p}, t_n) 
  &\approx&
  C_{sos}({\bf r}_{p}, 0)  - 
  \delta {\bf r}(t_n) \cdot \nabla C_{sos}({\bf r}_{p}, 0)
  \label{eq_corr_102}
\end{eqnarray}
where we have assumed that ${\mathcal A}^{-1}(0) = {\mathcal I}$, the identity
operator.

\vspace{20pt}

\bibliographystyle{elsarticle-harv}
\bibliography{mag_res}

\begin{thebibliography}{28}
\expandafter\ifx\csname natexlab\endcsname\relax\def\natexlab#1{#1}\fi
\expandafter\ifx\csname url\endcsname\relax
  \def\url#1{\texttt{#1}}\fi
\expandafter\ifx\csname urlprefix\endcsname\relax\def\urlprefix{URL }\fi

\bibitem[{Arfken(1985)}]{arfken}
Arfken, G., 1985. Mathematical Methods for Physicists. Academic Press, New
  York.

\bibitem[{Bannister et~al.(2007)Bannister, Brady, and Jenkinson}]{Bannister}
Bannister, P.~R., Brady, J.~M., Jenkinson, M., 2007. Integrating temporal
  information with a non-rigid method of motion correction for functional
  magnetic resonance images. Image and Vision Computing 25, 311--320.

\bibitem[{Cox(1996)}]{AFNI}
Cox, R.~W., 1996. Afni: Software for analysis and visualization of functional
  magnetic resonance neuroimages. Comput. Biomed. Res. 29, 162--173.

\bibitem[{Friston et~al.(1995)Friston, Ashburner, Frith, Poline, and
  Heather}]{SPM2}
Friston, K.~J., Ashburner, J., Frith, C.~D., Poline, J.~B., Heather, J.~D.,
  1995. Spatial normalization and registration of images. Hum. Brain Mapp. 3,
  165--189.

\bibitem[{Giovannetti et~al.(2008)Giovannetti, Viti, Liu, Yu, Mittra, Landini,
  and Benassi}]{coil_load}
Giovannetti, G., Viti, V., Liu, Y., Yu, W., Mittra, R., Landini, L., Benassi,
  A., 2008. An accurate simulator for magnetic resonance coil sensitivity
  estimation. Concepts Magn. Reson. 33B, 209--215.

\bibitem[{Griswold et~al.(2002)Griswold, Jakob, Heidemann, Nittka, Jellus,
  Wang, Kiefer, and Haase}]{griswold1}
Griswold, M.~A., Jakob, P.~M., Heidemann, R.~M., Nittka, M., Jellus, V., Wang,
  J., Kiefer, B., Haase, A., 2002. Generalized autocalibrating partially
  parallel acquisitions (grappa). Magn. Reson. Med. 47, 1202--1210.

\bibitem[{Hajnal et~al.(1994)Hajnal, Myers, Oatridge, Schwieso, Young, and
  Bydder}]{Hajnal}
Hajnal, J.~V., Myers, R., Oatridge, A., Schwieso, J.~E., Young, I.~R., Bydder,
  G.~M., 1994. Artifacts due to stimulus correlated motion in functional
  imaging of the brain. Magn. Reson. Med. 31, 283--291.

\bibitem[{Hartwig et~al.(2011)Hartwig, Engstrom, Flodmark, Ingvar, and
  Skare}]{Hartwig}
Hartwig, A., Engstrom, M., Flodmark, O., Ingvar, M., Skare, S., 2011. A simple
  method to reduce signal fluctuations in fmri caused by the interaction
  between motion and coil sensitivities. Proc. Intl. Soc. Mag. Reson. Med. 19,
  3628.

\bibitem[{Jenkinson et~al.(2002)Jenkinson, Bannister, Brady, and
  Smith}]{MCFLIRT}
Jenkinson, M., Bannister, P.~R., Brady, J.~M., Smith, S.~M., 2002. Improved
  optimisation for the robust and accurate linear registration and motion
  correction of brain images. NeuroImage 17, 825--841.

\bibitem[{Jezzard and Balaban(1995)}]{JB}
Jezzard, P., Balaban, R.~S., 1995. Correction for geometric distortion in echo
  planar images from b0 field variations. Magn. Reson. Med. 34, 65--73.

\bibitem[{Jin(1989)}]{Jin}
Jin, J., 1989. Electromagnetic Analysis and Design in Magentic Resonance
  Imaging. CRC Press, 1999.

\bibitem[{Kaza et~al.(2009)Kaza, Klose, and Lotze}]{Kaza}
Kaza, E., Klose, U., Lotze, M., 2009. Comparison of bold-signal magnitude
  between a 32-channel and a 12-channel head coil. IFMBE Proceedings 25,
  213--216.

\bibitem[{Kaza et~al.(2011)Kaza, Klose, and Lotze}]{Kaza2}
Kaza, E., Klose, U., Lotze, M., 2011. Comparison of a 32-channel with a
  12-channel head coil: are there relevant improvements for functional imaging?
  J Magn Reson Imaging 34, 173--183.

\bibitem[{Kundu et~al.(2012)Kundu, Inati, Evans, Luh, and Bandettini}]{ICA}
Kundu, P., Inati, S.~J., Evans, J.~W., Luh, W.-M., Bandettini, P.~A., 2012.
  Differentiating bold and non-bold signals in fmri time series using
  multi-echo epi. NeuroImage 60, 1759–1770.

\bibitem[{Li et~al.(2009)Li, Wang, and Wang}]{Li}
Li, J., Wang, L., Wang, Y., 2009. Visual bold-fmri with 32 channel phased-array
  coil at 3.0{T} mri system: comparison with 12 channel coil. Proc. Intl. Soc.
  Mag. Reson. Med. 17, 1614.

\bibitem[{Muresan et~al.(2005)Muresan, Renken, Roerdink, and
  Duifhuis}]{Spin_hist}
Muresan, L., Renken, R., Roerdink, J. B. T.~M., Duifhuis, H., 2005. Automated
  correction of spin-history related motion artefacts in fmri: Simulated and
  phantom data. IEEE Trans. Biomed. Eng. 52, 1450--1460.

\bibitem[{Ogawa et~al.(1990)Ogawa, Lee, Nayak, and Glynn}]{bold}
Ogawa, S., Lee, T.~M., Nayak, A.~S., Glynn, P., 1990. Oxygenation-sensitive
  contrast in magnetic resonance image of rodent brain at high magnetic fields.
  Magn Reson Med 14, 68--78.

\bibitem[{Power et~al.(2011)Power, Barnes, Snyder, Schlaggar, and
  Petersen}]{Power}
Power, J.~D., Barnes, K.~A., Snyder, A.~Z., Schlaggar, B.~L., Petersen, S.~E.,
  2011. Spurious but systematic correlations in functional connectivity mri
  networks arise from subject motion. Neuroimage 59, 2142--2154.

\bibitem[{Pruessmann et~al.(1999)Pruessmann, Weiger, Scheidegger, and
  Boesiger}]{SENSE}
Pruessmann, K.~P., Weiger, M., Scheidegger, M.~B., Boesiger, P., 1999. Sense:
  Sensitivity encoding for fast mri. Magn. Reson. Med. 42, 952--962.

\bibitem[{Reykowski(2006)}]{Reykowski}
Reykowski, A., 2006. Design of dedicated mri systems for parallel imaging. In:
  Schönberg, S.~O., Dietrich, O., Reiser, M.~F. (Eds.), Parallel Imaging in
  Clinical MR Applications. Springer, pp. 155--159.

\bibitem[{Roche(2011)}]{Roche}
Roche, 2011. A four-dimensional registration algorithm with application to
  joint correction of motion and slice timing in fmri. IEEE Trans. Med. Imag.
  30, 1546--1554.

\bibitem[{Roemer et~al.(1990)Roemer, Edelstein, Hayes, Souza, Muellera, and
  Harris}]{Roemer}
Roemer, P.~B., Edelstein, A., Hayes, E., Souza, P., Muellera, .~M., Harris,
  R.~L., 1990. The nmr phased array. Magn. Reson. Med. 16, 192--225.

\bibitem[{Sheltraw et~al.(2012)Sheltraw, Deshpande, Trumpis, and
  Inglis}]{ACS_paper_2}
Sheltraw, D., Deshpande, V., Trumpis, M., Inglis, B., 2012. Simultaneous
  reduction of two common autocalibration errors in grappa epi time series
  data. ArXiv:1208.0972.

\bibitem[{Smith et~al.(2001)Smith, Bannister, Beckmann, Brady, Clare, Flitney,
  Hansen, Jenkinson, Leibovici, Ripley, Woolrich, and Zhang}]{FSL}
Smith, S., Bannister, P., Beckmann, C., Brady, M., Clare, S., Flitney, D.,
  Hansen, P., Jenkinson, M., Leibovici, D., Ripley, B., Woolrich, M., Zhang,
  Y., 2001. Fsl: New tools for functional and structural brain image analysis.
  Seventh Int. Conf. on Functional Mapping of the Human Brain.

\bibitem[{Van~Dijk et~al.(2012)Van~Dijk, Sabuncub, and Buckner}]{Dijka}
Van~Dijk, K. R.~A., Sabuncub, M.~R., Buckner, R.~L., 2012. The influence of
  head motion on intrinsic functional connectivity mri. Neuroimage 59,
  431--438.

\bibitem[{Walsh et~al.(2000)Walsh, Gmitro, and Marcellin}]{Ad_comb}
Walsh, D.~O., Gmitro, A.~F., Marcellin, M.~W., 2000. Adaptive reconstruction of
  phased array mr imagery. Magn. Reson. Med. 43, 682--690.

\bibitem[{Wiesinger et~al.(2006)Wiesinger, de~Moortele, Adriany, Zanche,
  Ugurbil, and Pruessmann}]{high_field_review}
Wiesinger, F., de~Moortele, P. F.~V., Adriany, G., Zanche, N.~D., Ugurbil, K.,
  Pruessmann, K.~P., 2006. Potential and feasibility of parallel mri at high
  field. NMR Biomed 19, 368--378.

\bibitem[{Woods et~al.(1991)Woods, Cherry, and Mazziotta}]{AIR}
Woods, R.~P., Cherry, S.~R., Mazziotta, J.~C., 1991. Rapid automated algorithm
  for aligning and reslicing pet images. J. Comput. Assist. Tomogr. 16,
  620--633.

\end{thebibliography}

\end{document}